\documentclass[prb,aps,showpacs,floatfix,superscriptaddress,twocolumn]{revtex4}

\usepackage{amsmath,amssymb,amsxtra,bm,bbm}
\usepackage[dvips]{graphicx}

\newcommand{\bp}{\textbf{p}}
\newcommand{\bq}{\textbf{q}}
\newcommand{\bk}{\textbf{k}}
\newcommand{\tg}{\tilde{g}}

\begin{document}

\title{1/N Expansion in  Correlated Graphene}

\author{Valeri N. Kotov}
\affiliation{Department of Physics, Boston University, 
590 Commonwealth Avenue, Boston, Massachusetts 02215}

\author{Bruno Uchoa}
\affiliation{Department of Physics, University of Illinois
 at Urbana-Champaign, Urbana, Illinois 61801}

\author{A. H. Castro Neto}
\affiliation{Department of Physics, Boston University, 
590 Commonwealth Avenue, Boston, Massachusetts 02215}


\begin{abstract}
\noindent
We examine the $1/N$ expansion, where $N$ is the number of two-component
Dirac fermions, for Coulomb interactions in graphene with a gap of 
magnitude $\Delta = 2 m$. We find that for $N\alpha\gg1$, where $\alpha$ is 
graphene's ``fine structure constant'', there is a crossover as a function
of distance $r$ from the usual 3D Coulomb law, $V(r) \sim 1/r$,  to a 
2D Coulomb interaction, $V(r) \sim \ln(N\alpha/mr)$, 
for $m^{-1} \ll r \ll m^{-1} N \alpha/6$. This effect reflects  
the weak ``confinement" of the electric field in the 
graphene plane. The crossover also leads to unusual renormalization 
of the quasiparticle velocity and gap at low momenta. We also discuss the 
differences between the interaction potential in gapped graphene and usual 
QED for different coupling regimes.
\newline
\noindent
\end{abstract}


\maketitle

\section{Introduction}

The physics of graphene, a two-dimensional (2D) allotrope of carbon, 
presents a unique opportunity to explore properties of gapless (massless) 
Dirac fermions in a solid-state context \cite{review}. Because graphene 
is a truly 2D material from electronic point of view, it is also an 
interesting system where one can explore the important issue of electron-electron 
interactions. The interaction problem is a central
 one for the physics of   low-dimensional  electron systems. 
Although the electron-electron interactions in graphene are not internally 
screened, that is, they remain long-range and decay like $1/r$ (the 
electric field lines propagate in 3D away from the graphene plane  
and one has the ordinary Coulomb law), there is little experimental evidence, 
if any, that electron-electron interactions play a role in graphene physics. 
It is possible that the interactions between the fermions are actually 
dielectrically screened by the presence of substrates, onto which 
graphene is deposited in most experimental setups. Nevertheless, there 
is a fast growing literature in suspended graphene samples \cite{suspended} 
that will eventually tell us much more about electron-electron interactions
in this amazing material.

In graphene, the strength of the Coulomb interaction relative to the 
kinetic energy is given by the dimensionless coupling constant (also
called graphene's ``fine structure constant''),
\begin{equation}
\alpha = \frac{e^2}{\hbar v},
\end{equation}
where $v$ is the Fermi velocity, and we absorbed the environmental
dielectric constant ($\epsilon_0$) into the definition of the charge
$e$. For graphene in vacuum, as in the case of suspended samples, 
$\alpha$ reaches its maximum value, $\alpha \approx 2.2$, i.e. 
interaction effects are expected to be strong. One of the ways 
strong-coupling effects can manifest themselves theoretically is 
through spontaneous generation of a mass $m$ (chiral symmetry breaking).
In solid state language, mass generation is equivalent to the opening
of a gap $\Delta=2|m|$ in the electronic spectrum. Hence, in this work, the 
terms ``mass'' and ``gap'' are interchangeable.
In relativistic quantum electrodynamics (QED) in two space (plus one time) 
dimensions, QED$_{2+1}$, the study  of this phenomenon started quite a while 
ago \cite{2dqed,criticalqed} and is still going strong today. Graphene is actually 
different from QED$_{2+1}$ because only the fermions are confined to a 2D 
plane, while the field lines extend through the whole 3D space. In addition, 
the Coulomb interaction in graphene can be considered instantaneous since
the speed of light $c$ is much larger than the Fermi velocity 
($v \approx c/300$). Hence, Lorenz invariance is not respected, 
which reflects the non-relativistic, purely band origin of the Dirac 
quasiparticles. For the case of graphene, mass generation has  been 
predicted \cite{kve,kve2}, although no experimental signatures have yet been 
detected. An in-plane magnetic field also favors an excitonic condensate (gap) \cite{tsvelik}.
There has  been a surge of recent numerical activity on
the problem of mass generation in graphene (without external field), \cite{drut} and a consensus 
seems to have been reached that above $\alpha_c \approx 1.1$, mass generation 
occurs \cite{kve,drut}.

Another way to analyze this phenomenon is as a function of  $N$, the number 
of fermion species, which for graphene is $N=4$ due to the spin (2) 
and valley (2) degeneracies.  In the strong coupling limit 
$\alpha \rightarrow \infty$, generation of mass occurs below a critical 
$N$ which was estimated to be $N_c \approx 7-9$ \cite{kve,drut}. 
In experiments, a detectable gap has so far been observed only in a 
situation when it is actually due to external factors, such as the 
presence of a substrate with specific symmetry, creating sublattice 
asymmetry in the graphene plane and thus making the graphene electrons 
massive (gapped) \cite{gap}. However, it is quite possible, as already 
mentioned, that in ``suspended" graphene, whose exploration has just 
begun, the gap generically exists due to the strong quasiparticle interaction.

Whether  graphene breaks overall parity (sublattice symmetry) in the 
process of spontaneous mass generation  depends on the details of the interactions. 
Long-range Coulomb interactions  \cite{kve} favor equally parity-even and parity-odd combinations
 of masses in the two Dirac cones (valleys).
 On the other hand in QED$_{2+1}$ it is usually argued that
 parity-breaking mass generation is not possible, but this has
 to do with the presence of vector interactions in the fully
relativistic model \cite{parity}.

In this work we study the effect of 
a finite gap on the quasiparticle interactions (in particular modifications 
 of Coulomb's law), and the renormalization of the quasiparticle parameters, 
such as the Fermi velocity and the gap itself.
 Our main goal has been to compute corrections to those quantities under the
 assumption that the system is already massive, e.g. due to external
 factors explicitly breaking the sublattice symmetry, as mentioned above. However at the end of
the paper we also present estimates how the new physics we find can
 possibly affect the spontaneous formation of a gap via the excitonic
 mechanism. We do not address the issue of parity breaking since
 only single valley Coulomb interactions are considered.

 For  massless graphene, 
the large-N limit was studied recently in Refs.~ [\onlinecite{son,foster}],
 extending earlier results \cite{Gonz}.  The present work can be viewed as 
an extension of those studies to the massive case. We also find that for  
massive Dirac fermions unconventional modification of the interaction 
vertex and  fermion's properties can occur, not possible for strictly massless 
quasiparticles. In particular, we find that there is a crossover regime
for $N \alpha\gg1$ where the 3D Coulomb law, $V(r) \sim 1/r$, is modified, due to the confinement of the electric field lines, to a 2D Coulomb law, $V(r) \sim \ln(1/r)$, 
with strong renormalizations of the quasiparticle properties. In this non-perturbative 
regime the photon field is confined to the graphene plane
leading to a situation where the Dirac electrons form a 2D ``relativistic'' Coulomb gas \cite{marino}. Such electronic state has never been observed
in nature before and, perhaps, with  developments in the control of the structure
of graphene samples, it can be studied soon.  

The paper is organized as follows. In the next section, for completeness, 
we analyze the weak 
coupling regime of small $\alpha$, which can also be relevant to real 
situations (when the substrate screening is strong). In section \ref{onen}
we study the $1/N$ expansion for the electronic properties of gapped
Dirac fermions and show the existence of this new intermediate regime
where weak confinement of the electric field lines leads to strong renormalization of
electron-electron interactions. We also calculate the implications
of this new regime in the quasiparticle properties.
Section \ref{conclusions} contains our conclusions, and in Appendix A some estimates
 related to the excitonic gap formation are summarized.

\section{Interaction potential: weak-coupling regime}
\label{weak}

Our starting point is a model of two-dimensional  massive Dirac fermions  with a gap 
$\Delta =2|m|$.  The  Hamiltonian of the system is \cite{review}

\begin{equation}
\label{ham1}
H = \sum_{\bf{p}} \Psi_{\bf{p}}^{\dagger} ( v {\bm \sigma} \cdot{\bf p} \pm m  \sigma_{3})
\Psi_{\bf{p}}  +  H_{I},
\end{equation}
where $H_{I}$ is the quasiparticle interaction
\begin{equation}
\label{ham2}
H_{I} = \frac{1}{2} \sum_{\bf{p}} n_{\bf p} V({\bf p})  n_{\bf-p}, \
\  n_{\bf p} \equiv \sum_{\bf{q}} \Psi_{\bf{q+p}}^{\dagger} \Psi_{\bf{q}},
\end{equation}
and the potential $V({\bf p})$ will be specified later.
 We work in a two-component representation, so that $\sigma_i, i=1,2,3$ are the Pauli
 matrices; $\hat{\sigma}_{0} = \hat{I}$ is the 2$\times$2 identity matrix, often omitted
 for simplicity, and the vector ${\bm \sigma}= (\sigma_1,\sigma_2)$.
Thus the Hamiltonian \eqref{ham1} describes the physics in a single Dirac cone (valley).
 The two valleys in graphene are connected by time reversal, which translates into
 opposite signs of the mass term: the sign $``+"$ in  \eqref{ham1}
 correspond to one of the valleys, while the sign $``-"$ to the other one.
All formulas that follow
 are invariant under a change of the mass sign. The valley (and spin) indexes in \eqref{ham1},\eqref{ham2}
are omitted for simplicity, and it is understood that the spinors $\Psi_{\bf{q}}$ and the density
$n_{\bf p}$ describe a given single valley.

 The low energy electronic spectrum (dispersion close
to the Fermi energy), for $H_{I}=0$, is given by:
\begin{equation}
\label{dis}
E^{\pm}_{\bf k} \ = \pm \sqrt{v^2{\bf k}^2 + m^2}.
\end{equation}
 The interaction $H_{I}$ renormalizes both $v$ and $m$, as we show
 later.
We will only analyze the case when the system is an insulator,
 i.e. the chemical potential is in the gap (e.g. fixed to be zero). 
It is quite remarkable that in recent experiments in gapped samples 
the chemical potential can actually be moved from the electron to the 
hole side through the gap, thus causing a metal-insulator transition 
\cite{gap2}. In addition to using $\hbar=1$ everywhere
we also put $v=1$ in all intermediate formulas and restore it 
only at the end, when necessary.

\begin{figure}[tb]
  \centering
  \includegraphics*[width=0.45\textwidth]{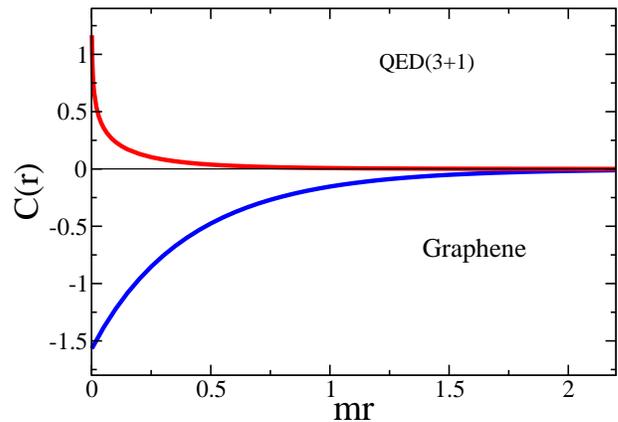}
  \caption{(Color online.) Plot of $C(r)$ as defined by   Eq.\eqref{pot0}, for the case of graphene (blue line) and
 3D QED (red line).}
  \label{Fig1}
\end{figure}

The polarization function, $\Pi({\bf q},\omega)$, for massive Dirac fermions
in the random phase approximation (RPA) was most recently analyzed in detail 
in Ref.~ [\onlinecite{vitor2}] (and can also be deduced by appropriate 
modification of the Lorentz-invariant results in QED$_{2+1}$ \cite{2dqed}). 
The result is:
\begin{equation}
  \label{pol3}
  \Pi(\bq,\omega) \! = \! -N\frac{|{\bf q}|^2}{4\pi} 
    \Biggl\{
      \frac{m}{q^2} + \frac{1}{2q}
        \biggl(1 \!-\!  \frac{4m^2}{q^2} \biggr) 
      \arctan{\Bigl(\frac{q}{2m}\Bigr)} 
  \Biggr\},
\end{equation}
where $q$ is the ``3-momentum'',
\begin{equation}
\label{q}
q=\sqrt{|{\bf q}|^2-\omega^2}.
\end{equation}
In RPA the effective potential is given by:
\begin{equation}
\label{charge}
V({\bf{q}})  =  \frac{V^{(0)}_{{\bf{q}}}}{1- V^{(0)}_{{\bf{q}}}\Pi(q)}, \ \ \
  V^{(0)}_{{\bf{q}}} = \frac{2\pi \alpha}{|{\bf{q}}|}
\end{equation}
Firstly, we study the behavior of the potential for weak coupling 
$N\alpha \ll 1$, and in this part of the work we fix $N$ to its 
graphene value $N=4$. The first order correction to the static potential 
is: $\delta V({\bf{q}}) \approx (V^{(0)}_{{\bf{q}}})^2 \Pi(\bq,\omega=0)$.
After transforming back to real space, we represent the correction by 
the function $C(r)$, and write the total potential as
\begin{equation}
  \label{pot0}
 V(r) \approx \frac{\alpha}{r} \Bigl(1 + \alpha  C(r)  + O(\alpha^2) \Bigr). \ \ \ 
\end{equation}
By using \eqref{pol3}, we obtain
\begin{eqnarray}
\label{Cr}
C(r) &= & -2(mr) \int_{0}^{\infty} dx J_{0}(2mrx)  \nonumber\\ 
&& \times\left \{ \frac{1}{x} + \biggl(1 -  \frac{1}{x^2} \biggr) 
      \arctan{(x)} \right \},
\end{eqnarray}
which is shown in Fig.~\ref{Fig1} (lower panel).
 
At this point we also find it useful to compare our results with the 
well studied case of conventional QED$_{3+1}$ \cite{BLP}, where 
interaction effects are also governed by a dimensionless coupling, 
the fine structure constant $\alpha_{QED}$. Eq.\eqref{pot0} has the same 
form (with the substitution $\alpha  \rightarrow \alpha_{QED} =1/137$),
leading to the so-called Uehling potential. The plot of $C_{QED}(r)$ in 
this case is shown in Fig.~\ref{Fig1} (upper panel) \cite{QED}.
It should be mentioned that (essentially as a consequence of the 
uncertainty principle) vacuum polarization effects are expected to be 
strong only at distances  below the Compton wavelength $r < \lambda_C = 1/m$. 
For our case we find, by evaluating the long distance asymptotic behavior 
of the integral \eqref{Cr} \cite{roman},
\begin{equation}
\label{ass}
C(r \gg m^{-1}) \sim -\sqrt{\frac{\pi}{mr}} \ e^{-2mr} \, .
\end{equation}
At short distances we find: $C(r \ll m^{-1}) = -\pi/2$, which 
is the well-known massless limit. 

Thus we observe two main differences 
between massive Dirac fermions and QED: (1) The sign of  the correction 
is different, i.e. in graphene vacuum polarization weakens the potential.
This can be traced to the fact that the charge itself $e^2$ is not 
renormalized in graphene \cite{vitor2}, while it diverges logarithmically 
at small distances in QED  (in other words the distribution of the vacuum 
polarization charge is very different in the two cases  \cite{vitor2}).
(2) The magnitude of the correction in graphene is more than 10 times 
larger (for typical distances), compared to QED (see Fig.~\ref{Fig1}). 
This seems to be related to the dimensionality of the problem. Thus, 
we conclude that (at weak coupling) polarization effects can be appreciable 
in a wide range of distances, especially if one assumes that Eq.\eqref{pot0} 
can be used also away from its strict applicability limit $\alpha \ll 1$. 
Typically such potential modification effects are important for calculation 
of localized energy levels (in the gap); so far such studies have not been 
performed experimentally in graphene.

The renormalization of $v$ and $m$ in the weak-coupling regime was addressed 
in Ref.~[\onlinecite{vitor2}] 
and will not be repeated here; the main difference from 
the massless case is that the mass provides an effective infrared cutoff 
where the renormalization stops, and the mass itself increases 
logarithmically (to first order in $\alpha$).

\section{$1/N$ expansion}
\label{onen}

Now, we analyze the limit $N\alpha \gg 1$, and proceed to evaluate
the renormalization of the potential and the quasiparticle properties.  
We view the calculation as a two-step procedure, which is not exact 
but will simplify the problem technically: first we calculate divergent 
terms that originate from intermediate integration in the high momentum 
region  ($m \ll |{\bf q}| \ll \Lambda$), where $\Lambda$ is the ultraviolet 
cutoff for graphene ($\Lambda \sim 1/a$, where $a \approx 1.42$ \, \AA \, 
is the lattice spacing) and after that, as a second step, we concentrate 
on renormalization from low momenta $|{\bf q}| < m$ (such renormalization 
will be more severe in the strong coupling limit $\alpha \to \infty$).

\subsection{High momentum regime: $m\ll|{\bf q}| \ll \Lambda$ }

To implement the first (large momentum) part of the scheme, one naturally 
expects the mass term to be unimportant; more formally, we can expand:
\begin{equation}
  \label{pol2}
  \Pi({\bf q}, \omega) = -\frac{N|{\bf q}|^2 }{4q} \left \{ \frac{1}{4} - \frac{m^2}{q^2} + ... \right \} \,,
  \quad 
  q \gg m
  \,.
\end{equation}
Keeping only the first term, one arrives at the effective potential, 
identical to the one found for the massless case \cite{Gonz,son}:
\begin{equation}
\label{potrpa}
V({\bf{q}}, \omega)  \approx \frac{2\pi \alpha}{|{\bf q}|}\left \{ 1 +
 \frac{\pi g}{8}\frac{|{\bf q}|}{q} \right \}^{-1} \, ,
\end{equation}
where we use the notation
\begin{equation}
g \equiv N \alpha,
\end{equation}
and $q$ is defined by Eq.\eqref{q}.

Now we evaluate the self-energy correction to one loop, in the limit
 $g \gg 1$, where the second term in  \eqref{potrpa}
 dominates. The self energy is proportional to $1/N$ in this case.
The Green's function is
\begin{equation}
  \label{GF}
  \hat{G}^{-1}(\bk,\omega)
 = \omega - v\,{\bm \sigma}\cdot\bk -
  m\, \sigma_3 - \hat{\Sigma}(\bk,\omega)+ i\eta {\mbox{sign}} (\omega)\ ,
\end{equation}
where the self-energy at one-loop level is
\begin{equation}
\label{se}
\hat{\Sigma}(\bk,\omega) = i \int \frac{d^{2}p d \varepsilon }{(2\pi)^{3}} 
\hat{G}_{0}(\bk +\bp,\omega +\varepsilon )
V(\bp, \varepsilon) \, .
\end{equation}
 Here $\hat{G}_{0}$ is the free Green's function.
By expanding
\begin{equation}
\hat{\Sigma} = \omega \Sigma_{0} + v{\bm \sigma}\cdot\bk \Sigma_{v} + m\sigma_3 \Sigma_{m}\, ,
\end{equation}
we find
\begin{equation}
\label{gf2}
\hat{G}({\bf k},\omega) =
 \frac{Z}{\omega  - Z(1+\Sigma_{v} )v{\bm \sigma}\cdot\bk -Zm(1+\Sigma_{m})\sigma_3}\, ,
\end{equation}
where $Z$ is the quasiparticle residue:
\begin{equation}
Z^{-1} = 1 - \Sigma_{0} \, .
\end{equation}

 The calculation of the velocity renormalization is then practically identical
 to the massless case \cite{son}, except one should keep in mind that
 the mass provides an effective infrared cutoff in  the integrals (due to the
 finite value of the dispersion at zero momentum \eqref{dis}). Thus we put the mass
 to zero  to avoid lengthy formulas and keep the above in mind.
 One finally obtains ($\delta v(k)$ stands for the velocity correction at one-loop order)
\begin{equation}
\frac{\delta v(k)}{v} = \Sigma_{0} + \Sigma_{v} \!= \!-i \frac{16}{N} \int_{k}^{\Lambda}\frac{ p dp}{(2\pi)}
\int_{-\infty}^{\infty}\frac{ d \omega}{(2\pi)}(p^2 \!-\!\omega^2)^{-3/2},
\end{equation}
 where $k$ is the external momentum (neglected in the Green's function), and written
 as an effective infrared cutoff (thus we assume $k \gg m$; in the opposite limit $m$ is the cutoff).
 By performing Wick's rotation $\omega \rightarrow i\omega$ (which avoids crossing any poles) \cite{BLP},
 one finds
\begin{equation}
\label{velocity1}
\delta v(k)/v = \frac{8}{\pi^2}\frac{1}{N} \ln(\Lambda/k), \ \ m \ll k \ll \Lambda  \, .
\end{equation}
As expected, the result is this region is identical to the massless case \cite{son}.

 For the mass renormalization we have, written in more detail,
\begin{eqnarray}
\label{mass}
\frac{\delta m(k)}{m} & =& \Sigma_{0} + \Sigma_{m} = -i \frac{16}{N} \int_{k}^{\Lambda}\frac{ p dp}{(2\pi)}
\int\frac{ d \omega}{(2\pi)} \frac{\sqrt{p^2-\omega^2}}{p^2} \nonumber\\
&& \times \left \{ \frac{p^2+\omega^2}{(p^2-\omega^2)^{2}} + \frac{1}{p^2-\omega^2} \right \} \, .
\end{eqnarray}
Here the first term in the curly brackets corresponds to $\Sigma_{0}$ and the second one 
 to $\Sigma_{m}$. The final result is
\begin{equation}
\label{mass1}
\delta m(k)/m = \frac{16}{\pi^2}\frac{1}{N} \ln(\Lambda/k), \ \ m \ll k \ll \Lambda \, .
\end{equation}

Finally, the quasiparticle residue $Z$ is determined by the behavior of $\Sigma_{0}$.
 However in the  limit $g \gg 1$ one encounters some complications, 
because even the frequency integral in the expression for  $\Sigma_{0}$ 
(see \eqref{mass}) is logarithmically divergent (although in the sum 
$\Sigma_{0} + \Sigma_{m}$ this additional divergence is not present).
This means that we have to use the full, finite $g$ potential from 
\eqref{potrpa}. Performing the calculation one finds
\begin{equation}
\Sigma_{0} = \frac{\alpha}{\pi} \ln(\Lambda/k) \int_{0}^{\infty}dx \frac{1-x^2}{(g\pi/8 + \sqrt{1+x^2})(1+x^2)^{3/2}} \, ,
\end{equation}
 and therefore,
\begin{equation}
Z \approx 1 - \frac{8}{\pi^2}\frac{1}{N}\ln(g\pi/4) \ln(\Lambda/k), \ g \gg 1, \  m \ll k \ll \Lambda \, .
\end{equation}

 The results for the mass and velocity renormalization \eqref{velocity1},
\eqref{mass1} can  be used to form renormalization group (RG) equations 
for these quantities. The reasoning is similar to the one presented, for example, in ref.[\onlinecite{son}] for the massless case. One integrates out the high momentum degrees of 
freedom, i.e. momentum  regions  $\Lambda > |\bp| > \Lambda_1 $, 
and the results vary with the quantity $\ln(\Lambda/\Lambda_1) \equiv l$.
As evident from \eqref{velocity1} and \eqref{mass1} the renormalization should
stop at a scale $\sim m$. For $m$ large enough and $N\alpha\gg1$, the functional form of the potential \eqref{potrpa} is not 
significantly affected by the RG flow before it stops. The RG equations in that case are: 
\begin{eqnarray}
\frac{dv}{dl}=\frac{8}{N\pi^2}v \, ,
\nonumber
\\
\frac{dm}{dl}=\frac{16}{N\pi^2}m \, ,
\end{eqnarray}
that have the solutions:
\begin{equation}
\label{renorm1}
m(k) = m \left (\frac{\Lambda}{k} \right)^{\frac{16}{N\pi^2}},  \  \
v(k) =  v \left (\frac{\Lambda}{k} \right)^{\frac{8}{N\pi^2}} \, ,
\end{equation}
which are valid in the region $m \ll k \ll \Lambda$. 
Here $m,v$ are the corresponding quantities at the ultraviolet scale 
$\Lambda$, i.e. their initial band values at the lattice scale.

\subsection{Low-momentum regime: $|{\bf q}|\ll m $ }

Now we proceed to analyze the low-momentum region. In this limit, 
the polarization \eqref{pol3} can be expanded to give:
\begin{equation}
  \label{pol1}
  \Pi({\bf q}, \omega) = -\frac{N|{\bf q}|^2 }{12\pi m} \left \{ 1 - \frac{q^2}{10m^2} + ... \right \} \,,
  \quad 
  q \ll m
  \,.
\end{equation}
We keep only the first term, as the other terms decrease quite fast in powers 
of $q/m$. Notice also that in this limit $ \Pi({\bf q}, \omega)$ becomes 
frequency independent. The corresponding RPA effective potential is:
\begin{equation}
  \label{rpa}
 V({\bf q}) 
\approx \frac{2\pi\alpha}{|{\bf q}|+  \tg |{\bf q}|^{2}/m}\,, \ \   |{\bf q}| \lesssim m \, ,
\end{equation}
where we have defined:
\begin{eqnarray}
\tg \equiv g/6 = N\alpha/6 \, .
\end{eqnarray}
By direct numerical evaluation of the polarization bubble \cite{vitor2}, 
we actually find that the above formula is valid even up to 
$|{\bf q}| \sim m$.

In the strict long-distance limit $|{\bf q}| \rightarrow 0$ the above 
potential tends to the pure Coulomb potential. However, in the limit 
$\tg \rightarrow \infty$ there is an intermediate  window of momenta, 
$m/\tg \ll |{\bf q}| < m$, where the potential crosses over to the 2D
 Coulomb's law, 
\begin{equation}
\label{rpa21}
V({\bf q}) \approx \frac{12 \pi m}{N} \frac{1}{|{\bf q}|^2}, \ \ m/\tg \ll |{\bf q}| < m.
\end{equation}
 In real space we have:
\begin{equation}
 \label{rpa2}
 V(r) \approx \frac{6}{N} m \ln{\Bigl (\frac{\tg}{mr} \Bigr )}\,, \ \  \frac{\tg}{m} \gg r \gg \frac{1}{m} \, .
\end{equation}

 We also comment  on some other situations in electrodynamics when the Coulomb
 potential can be strongly modified. 
 It is known that in compact QED$_{2+1}$ linear confinement can occur due 
to non-perturbative instanton effects \cite{Polyakov}. It is also possible to 
have a confining potential in ferroelectrics where compact field configurations are favored leading to linear confinement of charges \cite{Kir}.
 Intermediate logarithmic behavior (in real space), similar to \eqref{rpa2}, 
can occur in thin films (of thickness $d$) with large dielectric constant 
$\kappa \gg 1$ \cite{boriss}. In that case the logarithmic behavior is
 limited to the region: $d \ll r \ll \kappa d$, i.e. $d$ plays the role 
(formally) of the ``Compton" wavelength $1/m$ in our case, and $ \kappa d$ 
plays the role of $N \alpha/m$. It was argued that such an intermediate 
regime can actually lead to modification of the variable-range-hopping law 
in systems with strong disorder. Finally, a mechanism similar to the one 
found in this work,  i.e. due to the dominance of fluctuations over the bare 
potential, was explored in the context of high temperature superconductivity 
models based on QED (where it contributes to spinon deconfinement-confinement 
transition) \cite{Spinons}.

\begin{figure}[tb]
  \centering
  \includegraphics*[width=0.46\textwidth]{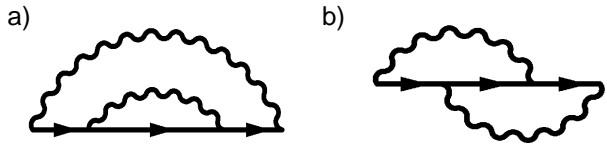}
  \caption{Two diagrams contributing to the self-energy at two-loop level.
 The wavy line stands for the potential \eqref{rpa}.}
  \label{Fig2}
\end{figure}

 Now we show that the low-momentum region where \eqref{rpa} is valid,
 also contributes singularly to the self-energy, in the  strong-coupling limit $\alpha \gg 1$.
 Using again the expression \eqref{se} with the potential \eqref{rpa},
 we obtain for the velocity renormalization at one-loop level, 
\begin{equation}
\frac{\delta v^{(1)}(k)}{v} =  i\int_{k}^{m}\frac{p dp}{(2\pi)}
\int\frac{ d \omega}{(2\pi)} \frac{1}{\omega^2-m^2}V(|\bp|)\ .
\label{first}
\end{equation}
 Here we  have put the fermion energies $E_{\bp}  \approx m$. Because 
the potential is static, there is no residue renormalization ($Z=1$) in 
this momentum range, and we have finally
\begin{eqnarray}
\label{vel2}
\frac{\delta v^{(1)}(k)}{v} &=&  \frac{3}{N}\ln{\left (\frac{1+ \tilde{g}}{1+\tilde{g} k/m} \right )} 
\approx \frac{3}{N}\ln(m/k), \nonumber \\
&& {\mbox{ \hspace{2.5cm}  }  }  m/\tilde{g} \ll k \ll m \, .
\label{vFirst}
\end{eqnarray}
Performing the same calculation for the mass, one easily finds
\begin{equation}
\delta m^{(1)}(k)/m = \delta v^{(1)}(k)/v \ .
\label{mFirst}
\end{equation}

Let us also examine the two-loop self-energy which is given by the two diagrams
 of Fig.~\ref{Fig2}. After some rather involved calculations we obtain the
 first, ``rainbow" contribution (Fig.~\ref{Fig2}(a))
\begin{equation}
\hat{\Sigma}^{(2)}_{a} = -\frac{1}{N^2} \Bigl ( \frac{9}{2} v{\bm \sigma}\cdot\bk  + \frac{9}{4} m\sigma_3 
\Bigr )\ln(m/k) \, ,
\end{equation}
in the low-momentum region $m/\tilde{g} \ll k \ll m$.
For the vertex correction to the self-energy, 
shown in  Fig.~\ref{Fig2}(b), we find:
\begin{equation}
\hat{\Sigma}^{(2)}_{b} = -\frac{1}{N^2} \bigl (\omega  + 3 v{\bm \sigma}\cdot\bk  + 3 m\sigma_3 \bigr )\ln(m/k) \ .
\end{equation}
 From here we find the corrections to  $v,m,Z$ at this order:
\begin{eqnarray}
\label{}
\frac{\delta v^{(2)}}{v} &=& -\frac{17}{2}\frac{1}{N^{2}}\ln(m/k)\ , 
\nonumber \\
\frac{\delta m^{(2)}}{m} &=& -\frac{25}{4}\frac{1}{N^{2}}\ln(m/k)\ ,  
\nonumber \\
Z^{(2)} &\approx& 1 -  \frac{1}{N^{2}}\ln(m/k)\ ,
\label{second}
\end{eqnarray}
valid for $ m/\tg \ll k \ll m$.

Observe that a single logarithm appears at two loops, meaning that we do not
have a conventional  renormalization group situation (piling up of leading logs) in this low-momentum
 region. This behavior is similar to the case of a static Coulomb potential
 in the massless case, where a single log appears up to second order of 
perturbation theory \cite{Misch}.

 Let $m^{*},v^{*}$ be the values of these quantities at the lowest scale 
$k \sim m/\tg \ll m$. Keeping only the dominant one loop contribution, 
we can estimate
\begin{equation}
\label{renorm3}
v^{*}/v =  m^{*}/m \approx  1 + \frac{3}{N}\ln(\tg) + O(1/N^2) \ .
\end{equation}
From equations \eqref{dis},\eqref{renorm3} one can also deduce 
the correction to the dispersion at low momenta:
\begin{equation}
 |E_k| \approx m^{*} + \frac{(v^{*})^2k^2}{2m^{*}} \approx \left ( 1 + \frac{3}{N}\ln(\tg)
 \right ) \left  ( m + \frac{v^2k^2}{2m} \right ), 
\end{equation}
for $k \sim m/\tg \ll m$.

On the validity of the above expansions, comparing Eq. \eqref{vFirst} and \eqref{second} 
we note that in two loop the $1/N$ expansion 
breaks down for $N\approx N_c = 17/6$, when the coefficient in front of the log vanishes and the trend in the renormalization of the velocity is reversed towards increasing $\alpha$. In the strong coupling limit, $\alpha \to \infty$, the velocity renormalizes to zero, reinforcing $\alpha$ to be large,  what indicates the possibility of an instability below the critical $N$. Although this analysis is not directly applicable to graphene, where $N=4$, it is similar in spirit to the prediction of a metal-insulator transition in massless graphene \cite{kve,drut}, which would take place around the critical value $\alpha_c\approx1$, where usual perturbation theory in the Coulomb potential breaks down.

In this regard  it is also important to explore how  the change in the potential, 
 leading to the modified shape \eqref{rpa21}, can affect the  spontaneous formation
 of a mass gap via the non-perturbative excitonic mechanism \cite{kve,kve2}.  The complete 
self-consistent  examination of this problem is well beyond the scope of this work \cite{criticalqcd};
 however some estimates are presented in Appendix A. We can conclude (see
Eq.~\eqref{excitonicgap}) that the  gap increases  as $\tg$ increases, 
\begin{equation}
m \approx \Lambda e^{-\pi N/4 + (3\pi/4) \ln(\tg) } = \Lambda  \tg^{(3\pi/4)} e^{-\pi N/4} \ ,
\end{equation}
although this increase in
 small in the perturbative $1/N$ regime when  $\ln(\tg)/N \ll 1$. On the other hand 
 when 
\begin{equation}
\label{newscale}
\frac{3}{N}\ln(\tg) \sim 1 \ ,
\end{equation}
 the gap enhancement is
 substantial. However this  strong-coupling regime in not accessible
 within the conventional large $N$ philosophy, where $\tg = N \alpha/6$ is kept
 fixed while $N\gg1$, so that the RPA self-consistent scheme is well justified. 
One can also hope that the results represent correctly the behavior for $N$ 
fixed at its physical value ($N=4$)  with $\tg$ large at strong coupling 
($\alpha \gg 1$), but of course in this case diagrams beyond RPA might be 
important and their influence remains unclear. Therefore we cannot make
 a definite conclusion how the system behaves under the condition $\frac{3}{N}\ln(\tg) \sim 1$,
although on the surface of things this criterion signifies a transition into a new
 low-energy regime which in itself deserves further study.

\section{Conclusions}
\label{conclusions}

 In conclusion, we have studied the behavior of the interaction potential 
and the quasiparticle properties  both in the weak-coupling 
$g = N \alpha \ll 1$ and the ``strong-coupling" $g = N \alpha \gg 1$ regimes. 
In the latter case we have found an unconventional regime where the potential 
crosses over from the usual 3D Coulomb potential to a 2D logarithmic behavior, 
as if the field lines were confined in the plane. This is due to the fact 
that vacuum fluctuations dominate over the bare potential, although they do 
so in a limited momentum range. This physics can also lead to unusual 
renormalization of physical quantities at distances well beyond the Compton 
wavelength $1/m$ (i.e. momenta $q \ll m$), up to distances of order 
$g/m \gg 1/m$. Such effects have not been studied, to the best of our 
knowledge, in conventional QED due to the smallness of  $\alpha_{QED}$.
 In our case both  the mass gap and the velocity ``keep running" (increase) 
up to the larger distance $g/m$. Since for graphene in vacuum $g \approx 9$, 
one can hope that this physics is observable.  On the other hand, 
the presence of various numerical coefficients makes this somewhat difficult,  
 e.g. in \eqref{rpa2} $\tg = g/6 = 1.5$ is probably too small to create a 
large-enough intermediate  energy window. Nevertheless, from purely 
theoretical perspective, the massive case exhibits much richer behavior 
compared to the gapless one.
 
\acknowledgments
We are grateful to R. Barankov, E. Fradkin, E. Marino, V. Pereira, and 
O. Sushkov for many stimulating discussions. We thank Roman Barankov 
for showing us the derivation of Eq.~(\ref{ass}).
This work was supported in part by the Office of Science, U.S.
 Department of Energy under Contract DE-FG02-91ER45439 through the
 Frederick Seitz Materials Research Laboratory at the University of
 Illinois. AHCN acknowledges the partial support of the U.S. Department of Energy under
the grant DE-FG02-08ER46512.

\appendix
\section{Enhancement of excitonic instability due  to Coulomb's law modification at strong coupling}
\label{excitoneq}

The equation for excitonic pairing leading to generation of mass $m$ 
 can be derived by determining the self-energy in Eq.~(\ref{se}) self-consistently,
 i.e. substituting $\hat{G}_{0} \rightarrow \hat{G}$  in that equation. Even if the ``bare" mass
 is initially zero, a finite (momentum-dependent) mass is then non-perturbatively generated, and obeys the
 equation:
\begin{equation}
m(0) = \frac{1}{2} \int \frac{d^{2}k}{(2\pi)^{2}}\frac{V(|{\bf k}|)m(k)}{\sqrt{k^{2}+m(k)^{2}}}.
\end{equation}
 In the gap equation vertex corrections are ignored (in the spirit of the large $N$ approach), 
 and the static limit of the RPA potential $V(|{\bf k}|$ is often used \cite{kve2}.
 Naturally, if on the right-hand side the mass $m(k)=m$ were the non-zero mass  already
 present in the Hamiltonian, then the above equation would simply be   the first perturbative
 correction to the mass, discussed in Section III.

 It is known that the momentum dependence of the mass cannot be ignored,
 and is important down to momenta  $k \sim  m(0)$ when the solution levels off.
 The full analysis results in the presence of a critical $N$, below which the solution
 exists \cite{kve2,criticalqed}. However, in order to estimate the influence of the potential
 \eqref{rpa21} on the gap, we use a simplified  approach  which misses the existence of a
critical coupling, but provides a  good qualitative estimate (and is similar to the initial
 attacks on the  problem of chiral symmetry breaking \cite{2dqed}). 
 This approach amounts to ignoring the momentum dependence of the mass, and we
 then obtain:
\begin{equation}
m \approx  \int_m^{\Lambda} \frac{kdk}{4\pi}\left(\frac{16}{Nk}\right)\frac{m}{|E^{\pm}_{\bf k}|}
+  \int_{m/\tg}^{m} \frac{kdk}{4\pi}\left(\frac{12 \pi m}{Nk^2}\right)\frac{m}{|E^{\pm}_{\bf k}|}.
\end{equation}
Here $E^{\pm}_{\bf k}$ is given by \eqref{dis},  and in our units $v=1$. The interaction potential in the above
 equation is taken in the  limit $g \gg 1 $, 
and we have taken into account the fact that the potential shape 
 depends on the momentum range, as determined by Eqs.~\eqref{potrpa},\eqref{rpa21}. 
From here we find the solution
\begin{equation}
\label{excitonicgap}
m \approx \Lambda e^{-\pi N/4 + (3\pi/4) \ln(\tg) } = \Lambda  \tg^{(3\pi/4)} e^{-\pi N/4}.
\end{equation}
 It is clear that in the regime $\ln(\tg)/N \ll 1$   we have an exponentially small solution.
 Only when $\frac{3}{N}\ln(\tg) \sim 1$,
 a substantial enhancement of the excitonic gap is possible, i.e. the exponential 
dependence 
 disappears, and the gap is proportional to  the large cutoff scale $\Lambda$.
 The condition $\frac{3}{N}\ln(\tg) \sim 1$  marks, as expected, a transition to a non-perturbative
 regime, where the perturbative $1/N$ corrections previously computed (see Eq.~\eqref{renorm3})
 become large, and the $1/N$ expansion breaks down.

\bibliographystyle{asprev}

\begin{thebibliography}{99}
\bibitem{QED} In conventional (3+1 dimensional) QED we have \cite{BLP}: \newline
 $C_{QED}(r) \!=\! 2/(3\pi) \! \int_{0}^{\infty} dx \ e^{-2mrx}\left [1+ (2x)^{-2} \right ]x^{-2}\sqrt{x^2-1}$,
\newline
 and $C_{QED}(r) \sim (4 \sqrt{\pi})^{-1} (mr)^{-3/2} \ e^{-2mr},  mr \gg 1$.
\end{thebibliography}

\end{document}